\documentclass[aps,prx,showpacs,twocolumn,superscriptaddress]{revtex4-1}
\usepackage{CJK}
\usepackage[tbtags]{amsmath}
\usepackage{amssymb}
\usepackage{bmpsize}
\usepackage{mathtools}
\usepackage{caption}

\usepackage{amsfonts}    
\usepackage{graphicx}   
\usepackage{verbatim}   
\usepackage{color}      
\usepackage{subfigure}  
\usepackage{hyperref}   
\usepackage{natbib}
\usepackage{booktabs}
\usepackage{threeparttable}
\usepackage{dcolumn}
\usepackage{multirow}
\usepackage{fancyhdr}

\begin{document}
\title{Quantum Information Transfer between a Two-Level and a Four-Level Quantum System}

\author{Tianfeng Feng}
\altaffiliation{These authors contributed equally}
\affiliation{State Key Laboratory of Optoelectronic Materials and Technologies and School of Physics, Sun Yat-sen University, Guangzhou, People's Republic of China}

\author{Qiao Xu}
\altaffiliation{These authors contributed equally}
\affiliation{State Key Laboratory of Optoelectronic Materials and Technologies and School of Physics, Sun Yat-sen University, Guangzhou, People's Republic of China}

\author{Linxiang Zhou}
\altaffiliation{These authors contributed equally}
\affiliation{State Key Laboratory of Optoelectronic Materials and Technologies and School of Physics, Sun Yat-sen University, Guangzhou, People's Republic of China}

\author{Maolin Luo}
\affiliation{State Key Laboratory of Optoelectronic Materials and Technologies and School of Physics, Sun Yat-sen University, Guangzhou, People's Republic of China}

\author{Wuhong Zhang}
\affiliation{State Key Laboratory of Optoelectronic Materials and Technologies and School of Physics, Sun Yat-sen University, Guangzhou, People's Republic of China}
\affiliation{Department of Physics, Jiujiang Research Institute and Collaborative Innovation Center for Optoelectronic Semiconductors and Efficient Devices, Xiamen University, Xiamen 361005, China}

\author{Xiaoqi Zhou}\email{zhouxq8@mail.sysu.edu.cn}
\affiliation{State Key Laboratory of Optoelectronic Materials and Technologies and School of Physics, Sun Yat-sen University, Guangzhou, People's Republic of China}

\begin{abstract}

Quantum mechanics provides a ``disembodied'' way to transfer quantum information from one quantum object to another. In theory, this quantum information transfer can occur between quantum objects of any dimension, yet the reported experiments of quantum information transfer to date have mainly focused on the cases where the quantum objects have the same dimension. Here we theoretically propose and experimentally demonstrate a scheme for quantum information transfer between quantum objects of different dimensions.
By using an optical qubit-ququart entangling gate, we observe the transfer of quantum information between two photons with different dimensions, including the flow of quantum information from a four-dimensional photon to a two-dimensional photon and vice versa.
The fidelities of the quantum information transfer range from $0.700$ to $0.917$, all above the classical limit of $2/3$. Our work sheds light on a new direction for quantum information transfer and demonstrates our ability to implement entangling operations beyond two-level quantum systems.

\end{abstract}


\maketitle

\begin{CJK}{UTF8}{gbsn}

\section{Introduction}

The information transfer between different objects is one of the most fundamental phenomena in nature.
In the classical world, a macroscopic object that carries unknown information can have its information precisely measured and copied, and thus this information can be transferred to another object while still being retained on the original object.
In the quantum world, although quantum mechanics does not allow the unknown quantum information carried on a quantum object to be perfectly cloned or precisely measured \cite{wootters, Gisin}, it does allow quantum information to be transferred from one object to another object in a disembodied way, i.e., only the quantum information but not the object itself is transferred. 
Quantum information transfer (QIT) between two quantum objects, which is also called quantum teleportation \cite{Bennett,Pirandola} when the two objects are separated at different locations, is widely used in quantum information applications including long-distance quantum communication \cite{Ekert,Ren,Ma}, distributed quantum networks \cite{Kimble,Simon} and measurement-based quantum computation \cite{Raussendorf,Raussendorf2,Walther, d-oneway,Gottesman}. It has been experimentally demonstrated in a variety of physical systems \cite{Kandel,He,Bouwmeester,Furusawa,Bao,Sherson,Riebe,Barrett,Qiao,Pfaff,Opt,Steffen,Yao}, including photons \cite{Bouwmeester,Furusawa}, atoms \cite{Bao}, ions \cite{Sherson,Riebe,Barrett}, electrons \cite{Qiao}, defects in solid states \cite{Pfaff}, optomechanical system \cite{Opt} and superconducting circuits \cite{Steffen, Yao}. Recently, more complex experiments have also been reported, such as the open destination teleportation \cite{Zhao, Barasinski} and the teleportation of a composite system \cite{Zhang,Luo}, a multi-level state \cite{Hu, Luo}, and multi-degree of freedom of a particle \cite{Wang}.

\begin{figure}[htbp]
	\includegraphics[width=0.45\textwidth]{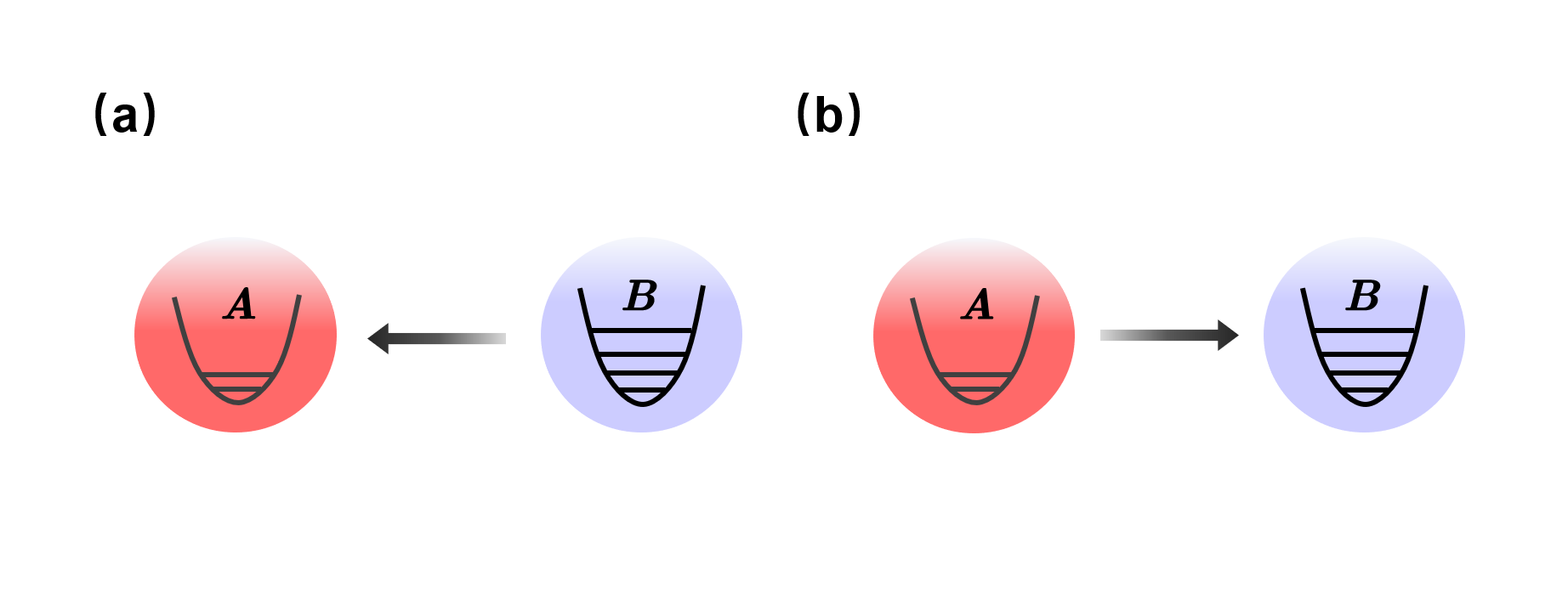}
	\caption{Quantum information transfer between a two-level and a four-level quantum system.  (a) Quantum information transfer from a four-level system B to a two-level system A. 
 (b) Quantum information transfer from a two-level system A to a four-level system B. 
 }
	\label{c}
\end{figure}

So far, the reported experiments have mainly focused on QIT between quantum objects with the same dimension. However, in quantum applications such as distributed quantum networks, different quantum objects may have different dimensions, and QIT between them is also required. 
For example, as shown in Fig.~1a, there are two quantum objects A and B, where A is two-dimensional and carries no quantum information, and B is four-dimensional and carries two qubits of unknown quantum information beforehand. If B wants to transfer quantum information to A, since A is two-dimensional and capable of carrying at most one qubit of quantum information, only one of the two qubits stored in B can be transferred to A.
We call this process a 4-to-2 QIT, which will distribute the two qubits of quantum information originally concentrated on B over both A and B. Now A and B are each loaded with one qubit of quantum information. Obviously, as shown in Fig.~1b, this one qubit of quantum information stored in A can also be transferred back to B. We call this process a 2-to-4 QIT, which concentrate the two qubits of quantum information distributed over both A and B on object B only.

In this work, we theoretically propose and experimentally demonstrate a scheme for QIT between a two-dimensional quantum object and a four-dimensional one.
By using an optical qubit-ququart entangling gate, we successfully transfer one qubit of quantum information from a four-dimensional photon preloaded with two qubits of quantum information to a two-dimensional photon, i.e., achieving a 4-to-2 QIT. 
We also experimentally realize a 2-to-4 QIT, i.e., transferring one qubit of quantum information from a two-dimensional photon to a four-dimensional photon preloaded with one qubit of quantum information.
Besides fundamental interests, the QITs demonstrated here have the potential to simplify the construction of quantum circuits and find applications in quantum computation and quantum communications.

\begin{figure}[htbp]
\includegraphics[width=0.49\textwidth]{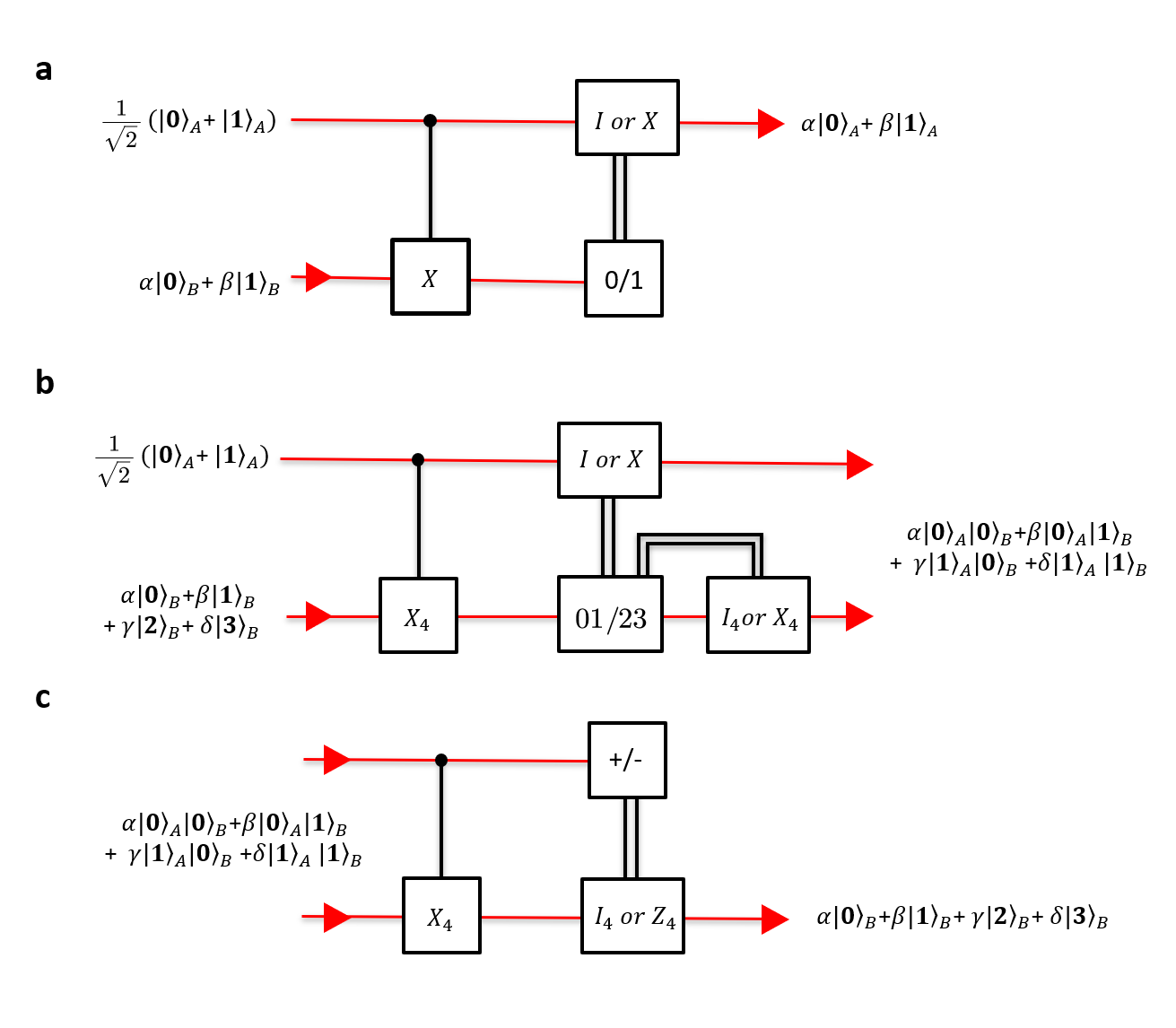}
\caption{
Schematic diagrams for quantum information transfer.
(a) The quantum information transfer from one qubit to another. The C$X$ gate entangles qubit A, which initially contains no quantum information, and qubit B, which initially contains one qubit of unknown quantum information. The projective measurement on B removes quantum information from B, thus transferring the one qubit of quantum information to A. After the feedforward unitary operation, the quantum information originally stored in B is restored in A, thus completing the quantum information transfer.
(b) The quantum information transfer from a ququart to a qubit. The initial state of ququart B contains two qubits of unknown quantum information while qubit A initially contains no quantum information. After entangling A and B using a C$X_4$ gate, where $X_4$ swaps $\vert 0\rangle$ and $\vert 2\rangle$ ($\vert 1\rangle$ and $\vert 3\rangle$), a projective measurement is applied on B to measure whether it is in the subspace spanned by $\vert 0\rangle$ and $\vert 1\rangle$ or the subspace spanned by $\vert 2\rangle$ and $\vert 3\rangle$. Based on the measurement result, feedforward unitary operations are applied on A and B, and the final state of A and B contains the two qubits of quantum information originally stored in B, thus completing the quantum information transfer from ququart B to qubit A.
(c) The quantum information transfer from a qubit to a ququart. Two qubits of unknown quantum information are initially distributed over qubit A and ququart B.
After applying a C$X_4$ gate on A and B, a projective measurement is applied on A and a feedforward unitary operation based on the measurement result is applied on B. 
The final state of B contains the two qubits of quantum information originally distributed over both A and B, thus completing the quantum information transfer from qubit A to ququart B.}
\label{scheme}
\end{figure}

\section{Scheme of Quantum Information Transfer}
\subsection{2-to-2 QIT}
Before presenting the scheme for implementing the QIT between quantum objects of different dimensions, let us briefly review how to implement the QIT between quantum objects of the same dimensions. 
Here we take the two-dimensional case as an example, assuming that both A and B are 2-dimensional quantum objects, where A is in the state $\frac{1}{\sqrt{2}}(\vert  \bold{0}\rangle_A+\vert  \bold{1}\rangle_A)$ without any quantum information preloaded, and B is in the state $\alpha \vert  \bold{0}\rangle_B+\beta \vert  \bold{1}\rangle_B$ loaded with one-qubit unknown quantum information.
As shown in Fig.~2a, by applying a controlled-X gate~(CX gate, commonly referred as CNOT gate) on A and B, the quantum state of the composite system of A and B would become
\[
\begin{split}
&\frac{1}{\sqrt{2}}\vert  \bold{0}\rangle_A(\alpha \vert  \bold{0}\rangle_B+\beta \vert  \bold{1}\rangle_B)+\frac{1}{\sqrt{2}}\vert  \bold{1}\rangle_A(\alpha \vert  \bold{1}\rangle_B+\beta \vert  \bold{0}\rangle_B)\\
=&\frac{1}{\sqrt{2}}(\alpha \vert  \bold{0}\rangle_A+\beta \vert  \bold{1}\rangle_A)\vert  \bold{0}\rangle_B+\frac{1}{\sqrt{2}}(\beta \vert  \bold{0}\rangle_A+\alpha \vert  \bold{1}\rangle_A)\vert  \bold{1}\rangle_B,
\end{split}
\]
where $X\vert \bold{0}\rangle=\vert \bold{1}\rangle$, $X\vert\bold{1}\rangle=\vert \bold{0}\rangle$.
After measuring  B in the $\vert \bold{0/1}\rangle$ basis and forwarding the measurement outcome to A, a unitary operation~($I$ or $X$) based on the outcome is applied on A and thus convert its state to $\alpha \vert  \bold{0}\rangle_A+\beta \vert  \bold{1}\rangle_A$. The state of A now has the same form as the initial state of B and thus completes the QIT from B to A.

\subsection{4-to-2 QIT}
We now present the scheme for realizing the QIT from a four-dimensional to a two-dimensional quantum object.
Suppose A is still two-dimensional and in the state $\frac{1}{\sqrt{2}}(\vert  \bold{0}\rangle_A+\vert  \bold{1}\rangle_A)$, and B is now four-dimensional and in the state  
\begin{equation}
\alpha \vert  \bold{0}\rangle_B+\beta \vert  \bold{1}\rangle_B+\gamma \vert  \bold{2}\rangle_B+\delta \vert  \bold{3}\rangle_B,
\end{equation} 
which is preloaded with two-qubit unknown quantum information.  To achieve the QIT between a two-dimensional and a four-dimensional quantum object, instead of using a two-qubit gate like the CNOT gate, one should use a qubit-ququart entangling gate.
As shown in Fig.~2b, a controlled-$X_4$ gate~(C$X_4$) is applied to A and B, where $X_4$ is a four-dimensional unitary gate defined as
\[
X_{4}=\left(      
      \begin{array}{cccc}
        0 & 0 & 1 & 0  \\
       0 & 0 & 0 & 1  \\
        1& 0 & 0 & 0\\
        0 & 1 & 0 & 0\\
      \end{array}
      \right),
\]
which converts $\vert \bold{0}\rangle$ ($\vert \bold{1}\rangle$) to $\vert \bold{2}\rangle$ ($\vert \bold{3}\rangle$) and vice versa. 
The state of  A and B is thus converted to
\[
\begin{split}
&\frac{1}{\sqrt{2}}\vert \bold{0}\rangle_A(\alpha\vert  \bold{0}\rangle_B+\beta \vert  \bold{1}\rangle_B+\gamma \vert  \bold{2}\rangle_B+\delta \vert  \bold{3}\rangle_B)\\
+&\frac{1}{\sqrt{2}}\vert \bold{1}\rangle_A(\alpha\vert  \bold{2}\rangle_B+\beta \vert  \bold{3}\rangle_B+\gamma \vert  \bold{0}\rangle_B+\delta \vert  \bold{1}\rangle_B)\\
=&\frac{1}{\sqrt{2}}(\alpha\vert  \bold{0}\rangle_A\vert  \bold{0}\rangle_B+\beta \vert  \bold{0}\rangle_A\vert  \bold{1}\rangle_B+\gamma \vert  \bold{1}\rangle_A\vert  \bold{0}\rangle_B+\delta \vert  \bold{1}\rangle_A\vert  \bold{1}\rangle_B)\\
+&\frac{1}{\sqrt{2}}(\alpha\vert  \bold{1}\rangle_A\vert  \bold{2}\rangle_B+\beta \vert  \bold{1}\rangle_A\vert  \bold{3}\rangle_B+\gamma \vert  \bold{0}\rangle_A\vert  \bold{2}\rangle_B+\delta \vert  \bold{0}\rangle_A\vert  \bold{3}\rangle_B).
\end{split}
\]
A projective measurement is then applied on B to measure whether it is in the subspace spanned by $\vert  \bold{0}\rangle_B$ and $\vert  \bold{1}\rangle_B$ or the subspace spanned by $\vert  \bold{2}\rangle_B$ and $\vert  \bold{3}\rangle_B$. Based on the measurement outcome, unitary operations ($I\otimes I_4$ or $X\otimes X_4$) are applied on A and B and their state becomes
\begin{equation}
\alpha\vert  \bold{0}\rangle_A\vert  \bold{0}\rangle_B+\beta \vert  \bold{0}\rangle_A\vert  \bold{1}\rangle_B+\gamma \vert  \bold{1}\rangle_A\vert  \bold{0}\rangle_B+\delta \vert  \bold{1}\rangle_A\vert  \bold{1}\rangle_B,
\end{equation}
where
\[
I_{4}=\left(      
      \begin{array}{cccc}
        1 & 0 & 0 & 0  \\
       0 & 1 & 0 & 0  \\
        0& 0 & 1 & 0\\
        0 & 0 & 0 & 1\\
      \end{array}
      \right).
\]
Comparing Eq. (2) with Eq. (1), it can be seen that the two quantum states have exactly the same form except for the difference in the state basis, which means that the two-qubit quantum information previously stored in B are now distributed over both A and B. In other words, one of the two qubits of quantum information originally stored in B is now transferred to A, thus achieving a 4-to-2 QIT.

\subsection{2-to-4 QIT}
We now show how the same quantum circuit can be used to implement a 4-to-2 QIT (the inverse of the above process), i.e., transferring one qubit of quantum information from a two-dimensional quantum object to a four-dimensional quantum object preloaded with one-qubit unknown quantum information.
The initial state of A and B can be written as
\begin{equation}
\alpha\vert  \bold{0}\rangle_A\vert  \bold{0}\rangle_B+\beta \vert  \bold{0}\rangle_A\vert  \bold{1}\rangle_B+\gamma \vert  \bold{1}\rangle_A\vert  \bold{0}\rangle_B+\delta \vert  \bold{1}\rangle_A\vert  \bold{1}\rangle_B,
\end{equation}
where both the two-dimensional A and the four-dimensional B are preloaded with one qubit of unknown quantum information. Note that the quantum state of A and B can be either an entangled state or a separable state. 
 As shown in Fig.~2c, a C$X_4$ gate is applied to A and B and their state is thus converted to
\[
\begin{split}
&\alpha\vert  \bold{0}\rangle_A\vert  \bold{0}\rangle_B+\beta \vert  \bold{0}\rangle_A\vert  \bold{1}\rangle_B+\gamma \vert  \bold{1}\rangle_A\vert  \bold{2}\rangle_B+\delta \vert  \bold{1}\rangle_A\vert  \bold{3}\rangle_B\\
=&\vert +\rangle_A(\alpha \vert  \bold{0}\rangle_B+\beta \vert  \bold{1}\rangle_B+\gamma \vert  \bold{2}\rangle_B+\delta \vert  \bold{3}\rangle_B)\\
+&\vert -\rangle_A(\alpha \vert  \bold{0}\rangle_B+\beta \vert  \bold{1}\rangle_B-\gamma \vert  \bold{2}\rangle_B-\delta \vert  \bold{3}\rangle_B),
\end{split}
\]
where $\vert\bold{\pm}\rangle=\frac{1}{\sqrt{2}}(\vert  \bold{0}\rangle\pm\vert  \bold{1}\rangle)$.
Similarly, after measuring A in the $\vert\pm\rangle$ basis and forwarding the outcome to B, a four-dimensional unitary operation ($I_4$ or $Z_4$) is applied on B conditioned on the outcome and thus converts its state to
\begin{equation}
\alpha \vert  \bold{0}\rangle_B+\beta \vert  \bold{1}\rangle_B+\gamma \vert  \bold{2}\rangle_B+\delta \vert  \bold{3}\rangle_B,
\end{equation}
where
\[
Z_{4}=\left(      
      \begin{array}{cccc}
        1 & 0 & 0 & 0  \\
       0 & 1 & 0 & 0  \\
        0& 0 & -1 & 0\\
        0 & 0 & 0 & -1\\
      \end{array}
      \right).
\]
Comparing Eq. (4) with Eq. (3), it can be seen that the two quantum states have exactly the same form except for the difference in the state basis, which means that the two qubits of quantum information previously distributed over both A and B are now concentrated in B. In other words, the one-qubit quantum information originally stored in A is now transferred to B, thus achieving a 2-to-4 QIT.


\begin{figure*}[htbp]
\includegraphics[width=\textwidth]{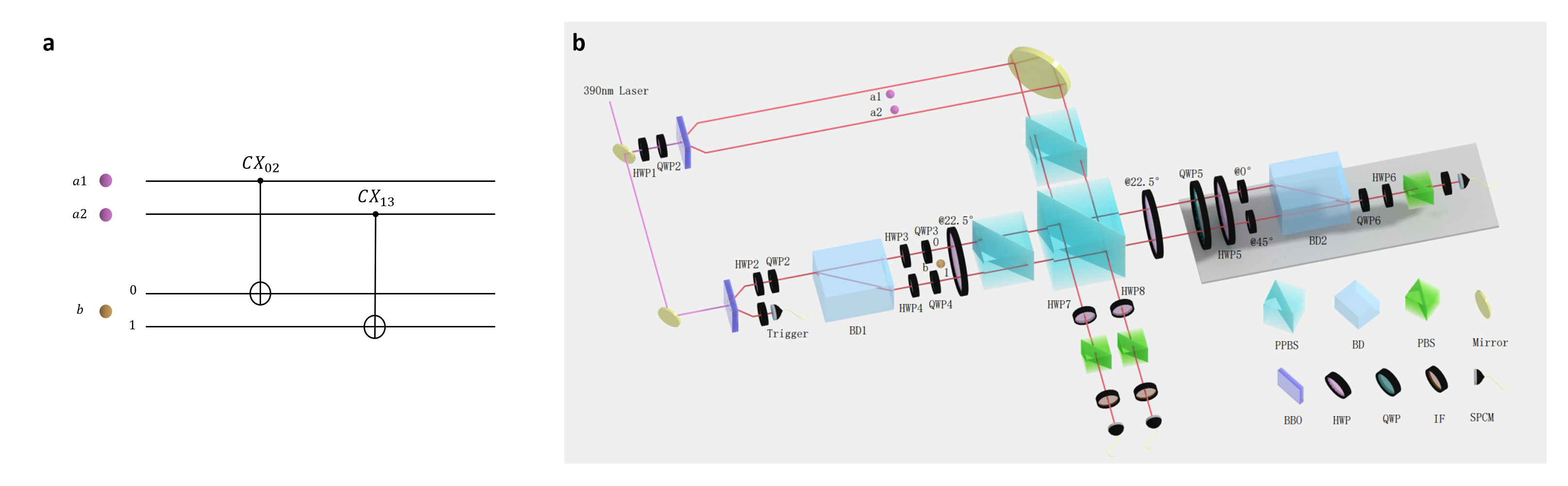}
\caption{
Experimental layout for quantum information transfer between a qubit and a ququart. 
(a) Optical C$X_4$ gate. Two photons a1 and a2 are used to encode qubit A, where $\vert  \bold{0}\rangle_A=\vert H\rangle_{a1}\vert H\rangle_{a2}$ and $\vert  \bold{1}\rangle_A=\vert V\rangle_{a1}\vert V\rangle_{a2}$. Photon b is used to encode ququart B, where $\vert \bold{0}\rangle_B=\vert H0\rangle_{b}$, $\vert \bold{1}\rangle_B=\vert H1\rangle_{b}$, $\vert \bold{2}\rangle_B=\vert V0\rangle_{b}$ and $\vert \bold{3}\rangle_B=\vert V1\rangle_{b}$. 
$H0$~($H1$) denotes photon in the upper~(lower) spatial mode with horizontal polarization and $V0$~($V1$)  denotes photon in upper~(lower) spatial mode with vertical polarization.
A C$X_{4}$ gate between the control qubit A and the target ququart B is decomposed into a C$X_{02}$ gate and a C$X_{13}$ gate. 
The C$X_{02}$ (C$X_{13}$) gate is equivalent to a polarization CNOT gate operating on photon a1 (a2) and photon b in the upper (lower) path.
(b) Experimental setup.
A pulsed ultraviolet (UV) laser is focused on two beta-barium borate (BBO) crystals and produces two photon pairs a1-a2 and b-t. By tuning HWP1 and QWP1, the first photon pair, a1-a2, is prepared at $\epsilon\vert H\rangle_{a1}\vert H\rangle_{a2}+\zeta\vert V\rangle_{a1}\vert V\rangle_{a2}$, which serves as the initial state of system A. BD1 and its surrounding waveplates (HWP2, QWP2, HWP3, QWP3, HWP4 and QWP4) prepare photon b at $\eta\vert H0\rangle_{b}+\kappa\vert H1\rangle_{b}+\lambda\vert V0\rangle_{b}+\mu\vert V1\rangle_{b}$, which serves as the initial state of system B. The two polarization CNOT gates based on PPBS are used to implement the optical C$X_4$ on system A and system B. BD2 and its surrounding waveplates (QWP5, HWP5, HWP$@0^{\circ}$, HWP$@45^\circ$, QWP6 and HWP6) are used to analyze the ququart state.
}
\label{setup}
\end{figure*}

\section{Experimental Demonstration using Linear Optics}
\subsection{Optical C$X_4$ gate}
To experimentally implement the QIT operations described above, the main challenge lies in the realization of the key part of the quantum circuit, namely the qubit-ququart entangling gate C$X_4$. 
Most experimentally realized quantum entangling gates so far are based on qubits~\cite{Nielsen,Barenco,Vidal,O'Brien,Chou,Fedorov,Reed}, and it is a challenging task to implement such high-dimensional entangling operations in any physical system. 

Here we present our method of implementing the C$X_4$ gate using linear optics. 
As shown in Fig.~3a, instead of implementing the C$X_4$ gate directly, we first decompose it into two consecutive gates C$X_{02}$ and C$X_{13}$ based on the fact that $X_4=X_{13}X_{02}$, where
\[
X_{02}=\left(
      \begin{array}{cccc}
        0 & 0 & 1 & 0  \\
       0 & 1 & 0 & 0  \\
        1& 0 & 0 & 0\\
        0 & 0 & 0 & 1\\
      \end{array}
      \right),~~~~ 
X_{13}=\left(
      \begin{array}{cccc}
        1 & 0 & 0 & 0  \\
       0 & 0 & 0 & 1  \\
        0& 0 & 1 & 0\\
        0 & 1 & 0 & 0\\
      \end{array}
      \right).
\]
Although $X_{02}$~($X_{13}$) is a four-dimensional unitary operation, it only operates on a two-dimensional subspace spanned by $\vert \bold{0}\rangle$ and $\vert \bold{2}\rangle$~($\vert \bold{1}\rangle$ and $\vert \bold{3}\rangle$).
 $X_{02}$~($X_{13}$) swaps $\vert \bold{0}\rangle$ and $\vert \bold{2}\rangle$~($\vert \bold{1}\rangle$ and $\vert \bold{3}\rangle$) and leaves $\vert \bold{1}\rangle$ and $\vert \bold{3}\rangle$~($\vert \bold{0}\rangle$ and $\vert \bold{2}\rangle$) unchanged. 
Based on this fact, two optical CNOT gates can be used to implement C$X_{02}$ and C$X_{13}$.

As shown in Fig.~3a, system A consists of two photons a1 and a2, serving as the control qubit, and system B consists of photon b, serving as the target ququart. The two orthonormal basis states of system A are $\vert  \bold{0}\rangle_A=\vert H\rangle_{a1}\vert H\rangle_{a2}$ and $\vert  \bold{1}\rangle_A=\vert V\rangle_{a1}\vert V\rangle_{a2}$, where $H$ and $V$ denote horizontal and vertical polarizations respectively. 
For system B, to encode a ququart with a single photon, photon b, both the polarization and spatial degree of freedoms are used, and the four orthonormal basis states are $\vert  \bold{0}\rangle_B=\vert H0\rangle_{b}$, $\vert  \bold{1}\rangle_B=\vert H1\rangle_{b}$, $\vert  \bold{2}\rangle_B=\vert V0\rangle_{b}$ and $\vert  \bold{3}\rangle_B=\vert V1\rangle_{b}$, where $H0$~($H1$) denotes photon in the upper~(lower) spatial mode with horizontal polarization and $V0$~($V1$)  denotes photon in the upper~(lower) spatial mode with vertical polarization. 
The C$X_{02}$ operation between qubit A and ququart B can be realized by applying a polarization CNOT gate to photon a1 and photon b in the upper path, which can be understood as follows: When the polarization of photon a1 is H (namely qubit A in $\vert  \bold{0}\rangle_A$), nothing happens; when the polarization of photon a is V (namely qubit A in $\vert  \bold{1}\rangle_A$), the polarization of photon b flips between H and V if b is in the upper path (namely ququart B's $\vert  \bold{0}\rangle_B$ and $\vert  \bold{2}\rangle_B$ components being swapped), and is unaffected if b is in the lower path (namely ququart B's $\vert  \bold{1}\rangle_B$ and $\vert  \bold{3}\rangle_B$ components being unchanged), which is exactly what a C$X_{02}$ gate achieves. Similarly, the C$X_{13}$ operation can be realized by applying a polarization CNOT gate to photon a2 and photon b in the lower path. As a result, the desired C$X_4$ gate can be achieved by the two polarization CNOT gates as shown in Fig.~3a.

Note that the use of two photons to encode the control qubit is mainly due to the following experimental considerations. The optical CNOT gates used experimentally are based on post-selection measurements, and such CNOT gates would fail when they act on the same two photons twice in a row. By using two photons a1 and a2 to encode the control qubit, the two CNOT gates are not acting on the same two photons, thus avoiding this problem.

\subsection{Experimental setup}
Figure 3b shows the experimental setup for implementing the two QIT schemes. Two photon pairs are generated by passing femtosecond-pulse UV laser through type-II beta-barium borate~(BBO) crystals~(See Appendix A: Generating two photon pairs). 
Photons a1 and a2 of the first pair are prepared at $\epsilon\vert H\rangle_{a1}\vert H\rangle_{a2}+\zeta\vert V\rangle_{a1}\vert V\rangle_{a2}$, which serves as the initial state of system A. 
By passing through the beam displacer BD1 and its surrounding waveplates (HWP2, QWP2, HWP3, QWP3, HWP4 and QWP4), the photon b from the second pair is prepared at $\eta\vert H0\rangle_{b}+\kappa\vert H1\rangle_{b}+\lambda\vert V0\rangle_{b}+\mu\vert V1\rangle_{b}$, which serves as the initial state of system B. 
The upper~(lower) rail of photon b is then superposed with photon a1~(a2) on a partial polarization beamsplitter~(PPBS). The PPBS, the loss elements and its surrounding half wave-plates can realize a polarization CNOT gate on photon a1~(a2) and photon b in the upper~(lower) path \cite{CNOT}. These two polarization CNOT gates together realize a C$X_4$ gate on qubit A and ququart B (See Appendix C: Implementation of C$X_4$ gate using linear optics), and the state of the three photons becomes
\[
\begin{split}
&\epsilon\vert H\rangle_{a1}\vert H\rangle_{a2}\otimes(\eta\vert H0\rangle_{b}+\kappa\vert H1\rangle_{b}+\lambda\vert V0\rangle_{b}+\mu\vert V1\rangle_{b})\\
+&\zeta\vert V\rangle_{a1}\vert V\rangle_{a2}\otimes(\eta\vert V0\rangle_{b}+\kappa\vert V1\rangle_{b}+\lambda\vert H0\rangle_{b}+\mu\vert H1\rangle_{b}).\\
\end{split}
\]

\subsection{Results of the 4-to-2 QIT}
To demonstrate the 4-to-2 QIT, $\epsilon$ and $\zeta$ are set to ${1}/{\sqrt{2}}$ and the three-photon state after the C$X_4$ gate can be written as
\[
\begin{split}
&\frac{1}{\sqrt{2}}\vert H\rangle_{a1}\vert H\rangle_{a2}\otimes(\eta\vert H0\rangle_{b}+\kappa\vert H1\rangle_{b}+\lambda\vert V0\rangle_{b}+\mu\vert V1\rangle_{b})\\
+&\frac{1}{\sqrt{2}}\vert V\rangle_{a1}\vert V\rangle_{a2}\otimes(\eta\vert V0\rangle_{b}+\kappa\vert V1\rangle_{b}+\lambda\vert H0\rangle_{b}+\mu\vert H1\rangle_{b}).\\
\end{split}
\]
Active feed-forward is needed for a full, deterministic 4-to-2 QIT. However, in this proof-of-principle experiment, we did not apply feed-forward but used post-selection to realize a probabilistic 4-to-2 QIT.
By post-selecting the $\vert H0\rangle_{b}$ and $\vert H1\rangle_{b}$ components  and converting $\vert H0\rangle_{b}$ to $\vert H\rangle_{b}$ and $\vert H1\rangle_{b}$ to $\vert V\rangle_{b}$ using BD2 and its preceding waveplates, the three-photon state
\[
\begin{split}
\eta\vert H\rangle_{a1}\vert H\rangle_{a2}\vert H\rangle_{b}+\kappa\vert H\rangle_{a1}\vert H\rangle_{a2}\vert V\rangle_{b}\\
+\lambda\vert V\rangle_{a1}\vert V\rangle_{a2}\vert H\rangle_{b}+\mu\vert V\rangle_{a1}\vert V\rangle_{a2}\vert V\rangle_{b}
\end{split}
\]
is obtained. By projecting photon a2 to $\vert D\rangle=\frac{1}{\sqrt{2}}(\vert H\rangle+\vert V\rangle)$, the two-photon state of a1 and b becomes
\[
\eta\vert H\rangle_{a1}\vert H\rangle_{b}+\kappa\vert H\rangle_{a1}\vert V\rangle_{b}
+\lambda\vert V\rangle_{a1}\vert H\rangle_{b}+\mu\vert V\rangle_{a1}\vert V\rangle_{b}.
\]
The two qubits of quantum information originally concentrated on photon b are now distributed over two photons a1 and b, which indicates that one qubit of quantum information has been transferred from photon b to photon a1, thus achieving a 4-to-2 QIT.

We then measure the fidelity of the final state, $F=$ Tr$(\rho|\psi\rangle\langle\psi|)$, which is defined as the overlap between the ideal final state ($|\psi\rangle$) and the measured density matrix ($\rho$). The verification of the QIT results is based on fourfold coincidence detection which in our experiment occur with a rate of 0.22 Hz. In each setting, the typical data collection time is 10 minutes, which allows us to sufficiently suppress Poisson noise.

Five different initial states of B are prepared for demonstrating the 4-to-2 QIT
\[
\begin{split}
&\vert \phi_{1}\rangle_B=\frac{1}{\sqrt{2}}(\vert  \bold{0}\rangle_B+\vert  \bold{1}\rangle_B)=\frac{1}{\sqrt{2}}(\vert H0\rangle_{b}+\vert H1\rangle_{b}),\\
&\vert \phi_{2}\rangle_B=\frac{1}{\sqrt{2}}(\vert  \bold{0}\rangle_B+\vert  \bold{2}\rangle_B)=\frac{1}{\sqrt{2}}(\vert H0\rangle_{b}+\vert V0\rangle_{b}),\\
&\vert \phi_{3}\rangle_B=\frac{1}{2}(\vert  \bold{0}\rangle_B+\vert  \bold{1}\rangle_B+\vert  \bold{2}\rangle_B+\vert  \bold{3}\rangle_B)\\
&=\frac{1}{2}(\vert H0\rangle_{b}+\vert H1\rangle_{b}+\vert V0\rangle_{b}+\vert V1\rangle_{b}),\\
&\vert \phi_{4}\rangle_B=\frac{1}{\sqrt{2}}(\vert  \bold{1}\rangle_B+\vert  \bold{2}\rangle_B)=\frac{1}{\sqrt{2}}(\vert H1\rangle_{b}+\vert V0\rangle_{b}),\\
&\vert \phi_{5}\rangle_B=\frac{1}{2}(\vert  \bold{0}\rangle_B-\vert  \bold{1}\rangle_B-\vert  \bold{2}\rangle_B-\vert  \bold{3}\rangle_B)\\
&=\frac{1}{2}(\vert H0\rangle_{b}-\vert H1\rangle_{b}-\vert V0\rangle_{b}-\vert V1\rangle_{b}).\\
\end{split}
\]
Figure 4a-e shows the 4-to-2 QIT results of the five different initial states on specific bases, from which the fidelities can be extracted. For each of the five initial ququart states $|\phi_1\rangle_B$ to $|\phi_5\rangle_B$, the fidelity of the final state of A and B is, in numerical sequence: $0.8860\pm0.0298$, $0.7686\pm0.0271$, $0.7342\pm0.0255$, $0.7375\pm0.0203$, and $0.8220\pm0.0164$, which are summarized in Fig.~4f. 

\begin{figure*}[htbp]
	\includegraphics[width=\textwidth]{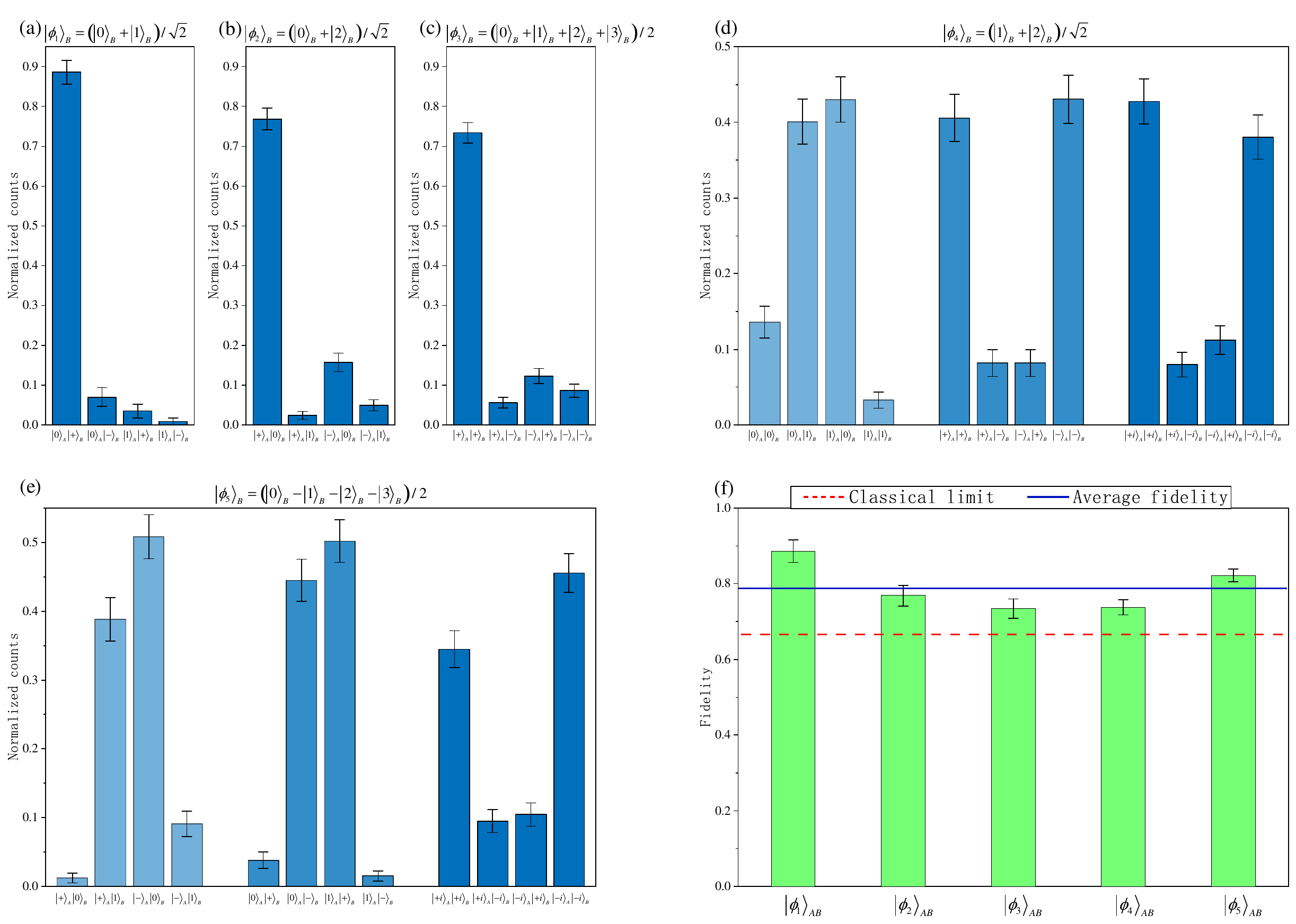}
	\caption{
		Experimental results for the quantum information transfer from ququart B to qubit A.
		(a-e) Measurement results of the final state of A and B for the initial states $|\phi_1\rangle_B$, $|\phi_2\rangle_B$, ... , and $|\phi_5\rangle_B$. Here 
		$|\pm\rangle=\frac{1}{\sqrt{2}}(|0\rangle \pm |1\rangle)$ and $|\pm i\rangle=\frac{1}{\sqrt{2}}(|0\rangle \pm i|1\rangle)$. 
		(f) Summary of the fidelities of the partial quantum state transfer for the five initial states.
		The average achieved fidelity of $0.7897\pm 0.0109$ overcomes the classical bound of 2/3. The error bars (SD) are calculated according to propagated Poissonian counting statistics of the raw detection events.
	}
	\label{setup}
\end{figure*}

\subsection{Results of the 2-to-4 QIT}
To demonstrate the 2-to-4 QIT, $\lambda$ and $\mu$ are set to zero and the quantum state after the C$X_4$ gate can be written as
\[
\begin{split}
&\epsilon\vert H\rangle_{a1}\vert H\rangle_{a2}\otimes(\eta\vert H0\rangle_{b}+\kappa\vert H1\rangle_{b})\\
+&\zeta\vert V\rangle_{a1}\vert V\rangle_{a2}\otimes(\eta\vert V0\rangle_{b}+\kappa\vert V1\rangle_{b}).
\end{split}
\]
By projecting both photons a1 and a2 to $\vert D\rangle$, the quantum state of photon b becomes
\[
\epsilon\eta\vert H0\rangle_{b}+\epsilon\kappa\vert H1\rangle_{b}+\zeta\eta\vert V0\rangle_{b}+\zeta\kappa\vert V1\rangle_{b}.
\]
This ququart state of photon b is then analyzed by the measurement setup consisting of BD2, its surrounding waveplates, the polarization beamsplitter (PBS) and the single-photon detector D3 (See Appendix D: State analysis of a photonic ququart state ).
The two qubits of quantum information originally distributed over system A (photons a1 and a2) and system B (photon b) is now concentrated on system B, which indicates that one qubit of quantum information has been transferred from system A to system B, thus achieving a 2-to-4 QIT.

\begin{figure*}[htbp]
	\includegraphics[width=\textwidth]{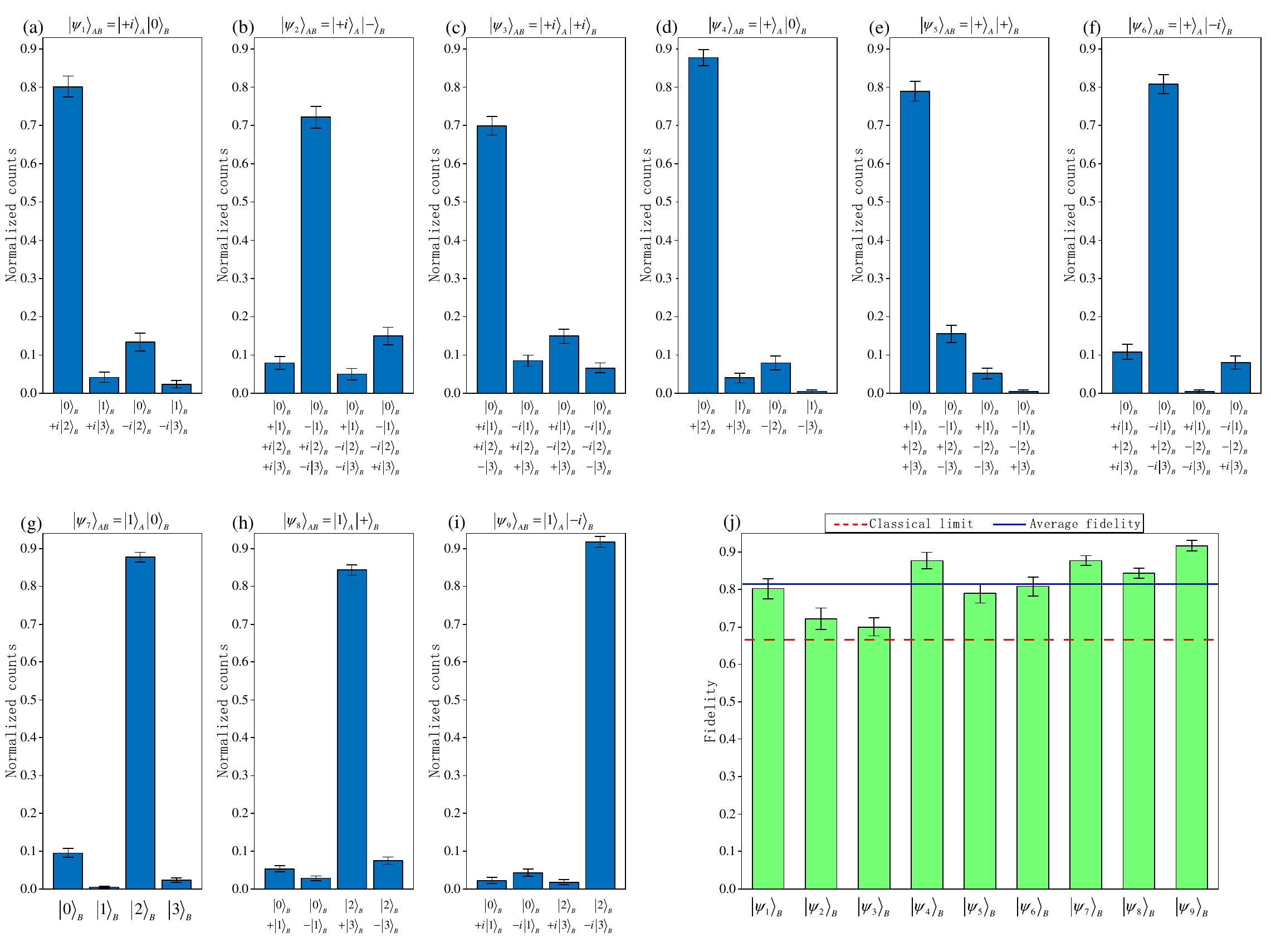}
	\caption{
		Experimental results for the quantum information transfer from qubit A to ququart B.
		(a-i) Measurement results of the final state of B for the initial states $|\psi_1\rangle_{AB}$, $|\psi_2\rangle_{AB}$, ... , and $|\psi_9\rangle_{AB}$. Here $|\pm\rangle=\frac{1}{\sqrt{2}}(|0\rangle \pm |1\rangle)$ and $|\pm i\rangle=\frac{1}{\sqrt{2}}(|0\rangle \pm i|1\rangle)$. 
		(j) Summary of the fidelities of the general quantum state transfer for the nine initial states.
		The average achieved fidelity of $0.8151\pm 0.0074$ overcomes the classical bound of 2/3. The error bars (SD) are calculated according to propagated Poissonian counting statistics of the raw detection events.
	}
	\label{setup}
\end{figure*}

Nine different initial states of A and B are used in the general quantum state transfer experiment
\[
\begin{split}
&\vert \psi_{1}\rangle_{AB}=\frac{1}{\sqrt{2}}(\vert  \bold{0}\rangle_A+i\vert  \bold{1}\rangle_A)\otimes\vert  \bold{0}\rangle_B\\
&=\frac{1}{\sqrt{2}}(\vert H\rangle_{a1}\vert H\rangle_{a2}+i\vert V\rangle_{a1}\vert V\rangle_{a2})\otimes\vert H0\rangle_{b},\\
&\vert \psi_{2}\rangle_{AB}=\frac{1}{\sqrt{2}}(\vert  \bold{0}\rangle_A+i\vert  \bold{1}\rangle_A)\otimes\frac{1}{\sqrt{2}}(\vert  \bold{0}\rangle_B-\vert  \bold{1}\rangle_B)\\
&=\frac{1}{\sqrt{2}}(\vert H\rangle_{a1}\vert H\rangle_{a2}+i\vert V\rangle_{a1}\vert V\rangle_{a2})\otimes\frac{1}{\sqrt{2}}(\vert H0\rangle_{b}-\vert H1\rangle_{b}),\\
\end{split}
\]
\[
\begin{split}
&\vert \psi_{3}\rangle_{AB}=\frac{1}{\sqrt{2}}(\vert  \bold{0}\rangle_A+i\vert  \bold{1}\rangle_A)\otimes\frac{1}{\sqrt{2}}(\vert  \bold{0}\rangle_B+i\vert  \bold{1}\rangle_B)\\
&=\frac{1}{\sqrt{2}}(\vert H\rangle_{a1}\vert H\rangle_{a2}+i\vert V\rangle_{a1}\vert V\rangle_{a2})\otimes\frac{1}{\sqrt{2}}(\vert H0\rangle_{b}+i\vert H1\rangle_{b}),\\
&\vert \psi_{4}\rangle_{AB}=\frac{1}{\sqrt{2}}(\vert  \bold{0}\rangle_A+\vert  \bold{1}\rangle_A)\otimes\vert  \bold{0}\rangle_B\\
&=\frac{1}{\sqrt{2}}(\vert H\rangle_{a1}\vert H\rangle_{a2}+\vert V\rangle_{a1}\vert V\rangle_{a2})\otimes\vert H0\rangle_{b},\\ 
&\vert \psi_{5}\rangle_{AB}=\frac{1}{\sqrt{2}}(\vert  \bold{0}\rangle_A+\vert  \bold{1}\rangle_A)\otimes\frac{1}{\sqrt{2}}(\vert  \bold{0}\rangle_B+\vert  \bold{1}\rangle_B)\\
&=\frac{1}{\sqrt{2}}(\vert H\rangle_{a1}\vert H\rangle_{a2}+\vert V\rangle_{a1}\vert V\rangle_{a2})\otimes\frac{1}{\sqrt{2}}(\vert H0\rangle_{b}+\vert H1\rangle_{b}),\\
&\vert \psi_{6}\rangle_{AB}=\frac{1}{\sqrt{2}}(\vert  \bold{0}\rangle_A+\vert  \bold{1}\rangle_A)\otimes\frac{1}{\sqrt{2}}(\vert  \bold{0}\rangle_B-i\vert  \bold{1}\rangle_B)\\
&=\frac{1}{\sqrt{2}}(\vert H\rangle_{a1}\vert H\rangle_{a2}+\vert V\rangle_{a1}\vert V\rangle_{a2})\otimes\frac{1}{\sqrt{2}}(\vert H0\rangle_{b}-i\vert H1\rangle_{b}),\\
&\vert \psi_{7}\rangle_{AB}=\vert  \bold{1}\rangle_A\otimes\vert  \bold{0}\rangle_B=\vert V\rangle_{a1}\vert V\rangle_{a2}\otimes\vert H0\rangle_{b},\\
\end{split}
\]
\[
\begin{split}
&\vert \psi_{8}\rangle_{AB}=\vert  \bold{1}\rangle_A\otimes\frac{1}{\sqrt{2}}(\vert  \bold{0}\rangle_B+\vert  \bold{1}\rangle_B)\\
&=\vert V\rangle_{a1}\vert V\rangle_{a2}\otimes\frac{1}{\sqrt{2}}(\vert H0\rangle_{b}+\vert H1\rangle_{b}),\\
&\vert \psi_{9}\rangle_{AB}=\vert  \bold{1}\rangle_A\otimes\frac{1}{\sqrt{2}}(\vert  \bold{0}\rangle_B-i\vert  \bold{1}\rangle_B)\\
&=\vert V\rangle_{a1}\vert V\rangle_{a2}\otimes\frac{1}{\sqrt{2}}(\vert H0\rangle_{b}-i\vert H1\rangle_{b}).\\
\end{split}
\]
Figure 5a-i shows the 2-to-4 QIT results of the nine different initial states on specific bases, from which the fidelities can be directly extracted. For each of the nine initial states $|\psi_1\rangle_{AB}$ to $|\psi_9\rangle_{AB}$, the fidelity of the final state of ququart B is, in numerical sequence: $0.8018\pm0.0271$, $0.7220\pm0.0289$, $0.6997\pm0.0241$, $0.8772\pm0.0217$, $0.7897\pm0.0257$, $0.8080\pm0.0249$, $0.8770\pm0.0130$, $0.8431\pm0.0134$, and $0.9171\pm0.0138$, which are summarized in Fig.~5j.

The reported data are raw data without any background subtraction, and the main error are due to double pair emission, imperfection in preparation of the initial states, and the non-ideal interference at the PPBS and BD2. 
Despite the experimental noise, the measured fidelities of the quantum states are all well above the classical limit ${2}/{3}$, defined as the optimal state-estimation fidelity on a single copy of a one-qubit system \cite{Massar}. These results prove the successful realization of the 4-to-2 and the 2-to-4 QIT.



	

\section{conclusion}
In this work, we have experimentally transferred one qubit of quantum information from a four-dimensional photon preloaded with two qubits of quantum information to a two-dimensional photon. We have also experimentally realized the inverse operation, namely transferring one qubit of quantum information from a two-dimensional photon to a four-dimensional photon preloaded with one qubit of quantum information. Our experiments show that quantum information is independent of its carriers and can be freely transferred between quantum objects of different dimensions. Although the present experiments are realized in the linear optical architecture, our protocols themselves are not limited to optical system and can be applied to other quantum systems such as trapped atoms \cite{Bao}, ions \cite{Riebe,Barrett}, and electrons \cite{Qiao}. 

The techniques developed in this work for entangling operations on photons of different dimensions can be used to prepare a new type of maximally entangled state such as $\frac{1}{2}(|0\rangle_{a1}|0\rangle_{a2}\otimes|0\rangle_{b}+|0\rangle_{a1}|1\rangle_{a2}\otimes|1\rangle_{b}+|1\rangle_{a1}|0\rangle_{a2}\otimes|2\rangle_{b}+|1\rangle_{a1}|1\rangle_{a2}\otimes|3\rangle_{b}) $, where photons a1 and a2 are both two-dimensional and belong to system A, and photon b is 4-dimensional and belongs to system B. System A and system B have the same dimension but different number of particles.
Such maximally entangled states with asymmetric particle numbers can be used as a physical resource for realizing quantum teleportation between systems with the same dimension but different particle numbers.

Our approach can be readily extended to higher dimensional cases (See Appendix E and F: Merge operation $\&$ Split operation). With these QIT operations, one can either concentrate the quantum information from multiple objects to one object or distribute the quantum information from one object to multiple objects. Such operations have the potential to simplify the construction of multi-qubit gates \cite{Ralph,Lanyon}
(See Appendix G: Construction of multi-qubit gates ) 
 and find applications in quantum computation and quantum simulations.

\end{CJK}



\section{Acknowledgements}
This work was supported by the National Key Research and Development Program of China (2017YFA0305200,  2016YFA0301300), the Key Research and Development Program of Guangdong Province (2018B030329001, 2018B030325001) and the National Natural Science Foundation of China (Grant No.61974168). T.F., Q.X. and L.Z. contributed equally.





\section{Apendixes}

\textbf{A. Generating two photon pairs.}
For the sake of simplicity, in Fig.~3b of the main text, we only show a simplified version of the spontaneous parametric down-conversion (SPDC) sources. Fig. 6 shows the detailed experimental setup for generating two photon pairs. An ultraviolet pulse laser centered at 390nm is split into two parts, which are used to generate two SPDC photon pairs. The lower part of the laser directly pumps a BBO crystal to generate a pair of photons in the state  $\left. \left. |V_b \right> |H_t \right> $ via beamlike type-II SPDC, where photon t is used for trigger. The upper part of the laser goes through HWP1 and QWP1 to prepare its polarization at $\alpha |\left. H \right> +\beta \left. |V \right> 
$. It then passes through an arrangement of two beam displacers (BDs) and HWPs to separate the laser into two beams by 4 mm apart (such configuration was first adopted by H.-S. Zhong et al. in \cite{ZhongHS}). The two beams then focus on a BBO crystal to generate two photon pairs in the states $\left. \left. |V_{a1} \right> |H_{a2} \right> $  and $\left. \left. |V_{a1'} \right> |H_{a2'} \right>$  via beamlike type-II SPDC, where the subscripts denote the spatial modes.   $\left. \left. |V_{a1} \right> |H_{a2} \right> $  and $\left. \left. |V_{a1'} \right> |H_{a2'} \right>$   are then rotated using HWPs to   $\left. \left. |H_{a1} \right> |H_{a2} \right> $  and $\left. \left. |V_{a1'} \right> |V_{a2'} \right>$  , respectively. Photon pairs of   $\left. \left. |H_{a1} \right> |H_{a2} \right> $  and $\left. \left. |V_{a1'} \right> |V_{a2'} \right>$   are then combined into same spatial modes using two BDs. After tilting the two BDs to finely tune the relative phase between the two components, the two photons a1 and a2 are prepared into $\alpha |\left. H \right> _{a1}|\left. H \right> _{a2}+\beta |\left. V \right> _{a1}|\left. V \right> _{a2}$, which is the desired quantum state of system A.

\begin{figure}
\includegraphics[width=0.5\textwidth]{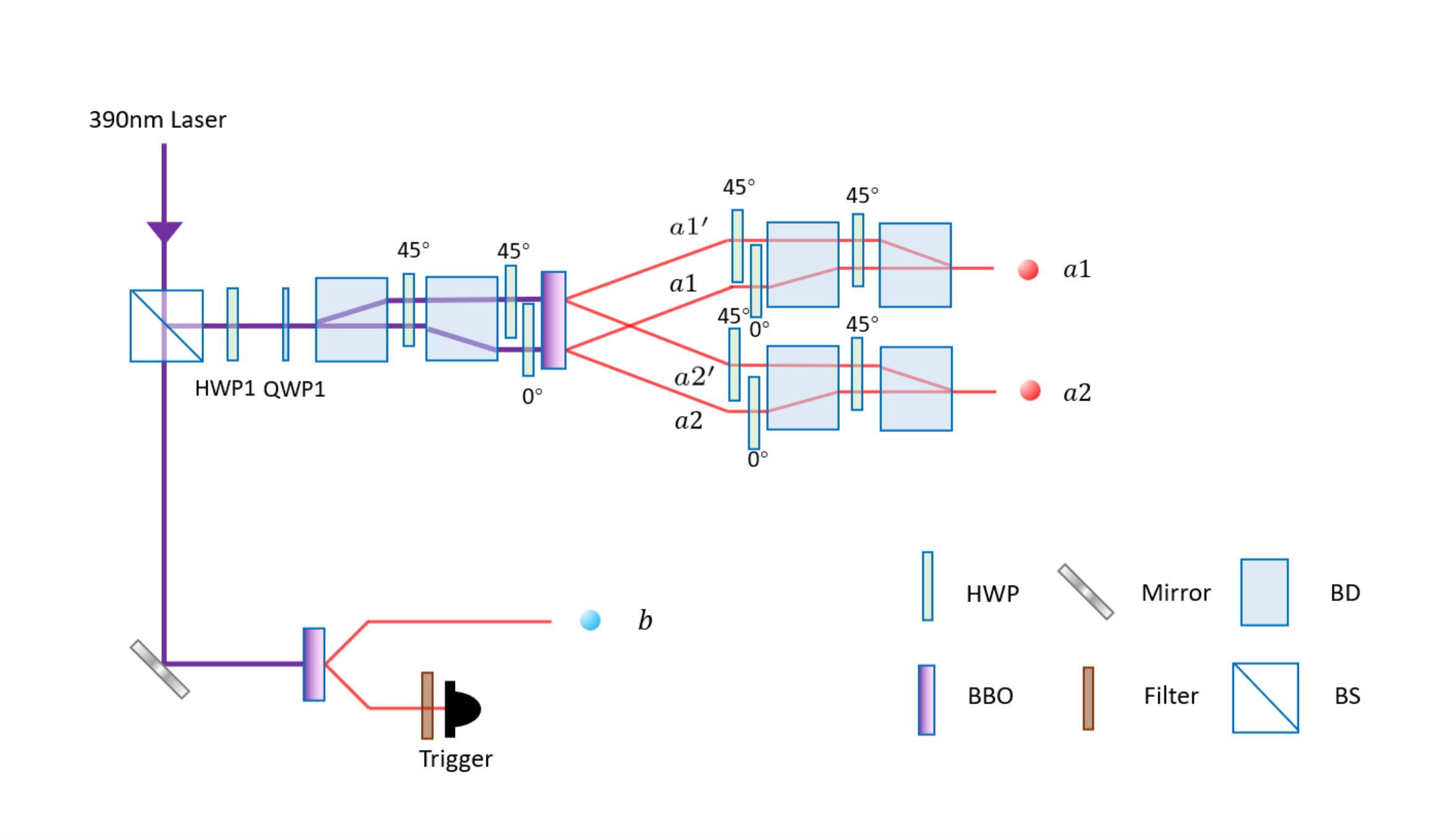}
\caption{Experimental setup for generating two photon pairs.}
\end{figure}

\textbf{B. Two-photon interference on a PPBS.}
The PPBS implements the quantum phase gate by reflecting vertically polarized light perfectly and reflecting (transmitting) $1/3$ ($2/3$) of horizontally polarized light. To realize a perfect quantum gate with the PPBS, the input photons on the PPBS need to be indistinguishable to each other. To evaluate the indistinguishability of the input photons, a two-photon Hong-Ou-Mandel (HOM) interference on the PPBS need to be measured. For large delay, the two photons are completely distinguishable due to their time of arrival. The probability to get a coincidence from a $|HH\rangle$ input is then $5/9$. In case of perfectly indistinguishable photons at zero delay, the probability drops to $1/9$. From the above considerations, the theoretical dip visibility $V_{th} =80\%$ is obtained, which is defined via $V=(c_\infty-c_0)/c_0$, where $c_0$ is the count rate at zero delay and $c_\infty$ is the count rate for large delay. As shown in Fig. 7, the HOM interference is experimentally measured and a dip visibility of $V_{exp} =0.661\pm0.0015$ is obtained, where the error bar is calculated from the Poissonian counting statistics of the detection events. The overlap quality $Q=V_{exp}/V_{th}=0.826\pm0.0019$ indicates that about $17.4\%$ of the detected photon pairs are distinguishable.

\begin{figure}
\includegraphics[width=0.4\textwidth]{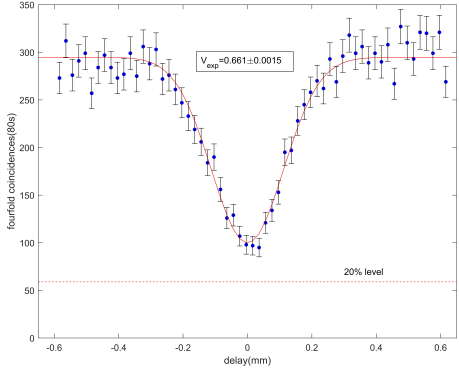}
\caption{HOM interference at the PPBS for a $|HH\rangle$ input. In case of perfect interference, the count rate should drop down to $20\%$, leading to a theoretically achievable dip visibility of $80\%$.
 }
\end{figure}



\textbf{C. Implementation of C$X_4$ gate using linear optics.}
As described in the main text, an optical C$X_4$ gate between system A (photons a1 and a2) and system B (photon b) can be implemented with a setup as shown in Fig.~8a. Photons a1 and a2 encode the control qubit  and its initial state is $\epsilon\vert  \bold{0}\rangle_A+\zeta\vert  \bold{1}\rangle_A=\epsilon\vert H\rangle_{a1}\vert H\rangle_{a2}+\zeta\vert V\rangle_{a1}\vert V\rangle_{a2}$. Photon b encodes the target ququart and its initial state is $\eta\vert  \bold{0}\rangle_B+\kappa\vert  \bold{1}\rangle_B=\eta\vert H0\rangle_{b}+\kappa\vert H1\rangle_{b}$.  After passing through the loss elements that transmit horizontally polarized light perfectly and transmit $1/3$ of vertically polarized light, photon a1 (a2) and photon b in the upper (lower) mode are superposed on the PPBS. The PPBS, the loss elements and the two surrounding HWPs at $22.5^\circ$ together implement a polarization CNOT operation on the input photons \cite{Langford, CNOT,Kiesel}. Such optical circuit can thus realize the following transformations
\[
\begin{split}
|H\rangle_{a1}|H\rangle_{a2}\otimes|H0\rangle_b&\longrightarrow|H\rangle_{a1}|H\rangle_{a2}\otimes|H0\rangle_b\\
|H\rangle_{a1}|H\rangle_{a2}\otimes|H1\rangle_b&\longrightarrow|H\rangle_{a1}|H\rangle_{a2}\otimes|H1\rangle_b\\
|V\rangle_{a1}|V\rangle_{a2}\otimes|H0\rangle_b&\longrightarrow|V\rangle_{a1}|V\rangle_{a2}\otimes|V0\rangle_b\\
|V\rangle_{a1}|V\rangle_{a2}\otimes|H1\rangle_b&\longrightarrow|V\rangle_{a1}|V\rangle_{a2}\otimes|V1\rangle_b.
\end{split}
\]
As a result, after passing through this optical circuit, the initial input state 
\[
\begin{split}
&(\epsilon\vert  \bold{0}\rangle_A+\zeta\vert  \bold{1}\rangle_A)\otimes (\eta|\bold{0}\rangle_{B}+\kappa|\bold{1}\rangle_{B})\\
=&(\epsilon|H\rangle_{a1}|H\rangle_{a2}+\zeta|V\rangle_{a1}|V\rangle_{a2})\otimes (\eta|H0\rangle_{b}+\kappa|H1\rangle_{b})
\end{split}
\]
 would become 
\[
\begin{split}
&\epsilon\vert H\rangle_{a1}\vert H\rangle_{a2}\otimes(\eta\vert H0\rangle_{b}+\kappa\vert H1\rangle_{b})\\
+&\zeta\vert V\rangle_{a1}\vert V\rangle_{a2}\otimes(\eta\vert V0\rangle_{b}+\kappa\vert V1\rangle_{b})\\
=&\epsilon\vert  \bold{0}\rangle_A\otimes(\eta|\bold{0}\rangle_{B}+\kappa|\bold{1}\rangle_{B})+\zeta\vert  \bold{1}\rangle_A\otimes(\eta|\bold{2}\rangle_{B}+\kappa|\bold{3}\rangle_{B}),
\end{split}
\]
which is exactly the desired output state of a C$X_4$ gate. The above optical C$X_4$ gate operates with a success probability of 1/27. In practice, to combat low count rates, we adopt the method proposed in reference \cite{CNOT} to simplify the implementation of the optical C$X_4$ gate. The simplified experimental setup is shown in Fig.~8b. We achieve correct balance by removing the loss elements and pre-biasing the input polarization states during gate characterization. The initial state of photons a1 and a2 is now prepared at $\frac{\epsilon}{\sqrt{|\epsilon|^2+\frac{|\zeta|^2}{9}}}\vert H\rangle_{a1}\vert H\rangle_{a2}+\frac{\zeta/3}{\sqrt{|\epsilon|^2+\frac{|\zeta|^2}{9}}}\vert V\rangle_{a1}\vert V\rangle_{a2}$ instead of $\epsilon|H\rangle_{a1}|H\rangle_{a2}+\zeta|V\rangle_{a1}|V\rangle_{a2}$. The HWP applied on photon b before entering the PPBS is now set at $15^\circ$, thus converting photon b to the state $\frac{\sqrt{3}}{2}(\eta|H0\rangle+\kappa|H1\rangle)+\frac{1}{2}(\eta|V0\rangle+\kappa|V1\rangle)$ instead of $\frac{1}{\sqrt{2}}(\eta|H0\rangle+\kappa|H1\rangle)+\frac{1}{\sqrt{2}}(\eta|V0\rangle+\kappa|V1\rangle)$, which is the case if the HWP is set at $22.5^\circ$.

\begin{figure}
	\includegraphics[width=0.45\textwidth]{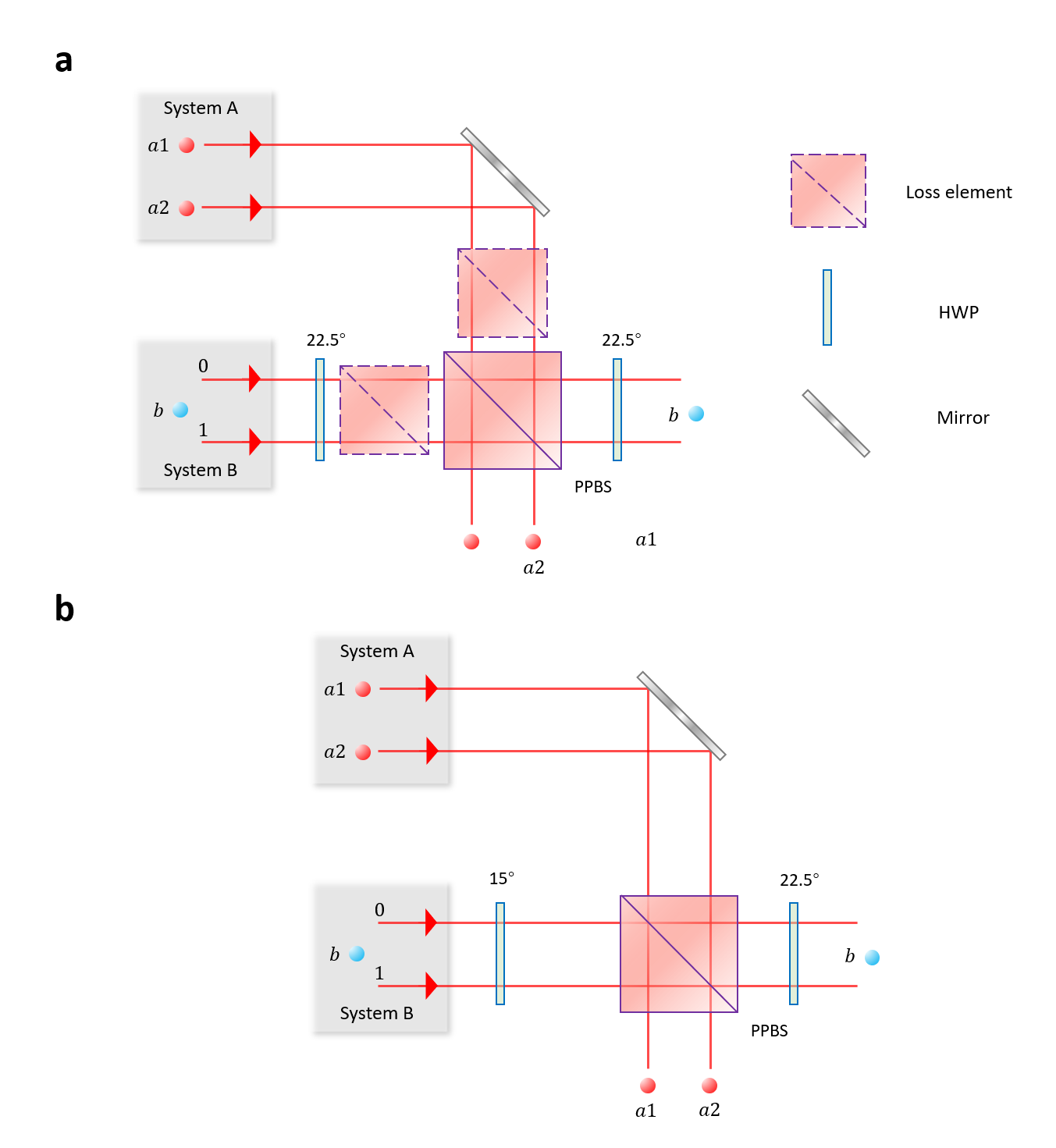}
	\caption{C$X_4$ gate with linear optics. $\bold{a,}$ The standard optical C$X_4$ gate. $\bold{b,}$ The simplified optical C$X_4$ gate.}\label{cx4}
\end{figure}

\textbf{D. State analysis of a photonic ququart state.}
A photon with both polarization and spatial degrees of freedom (DOFs) can encode a ququart state. To fully characterize such state, one needs to perform projective measurement onto various different ququart states. To fulfill this task, we build a ququart state analyzer as shown in Fig. 9, which can project the input ququart to any state in the form of $(a|H\rangle+b|V\rangle)\otimes(c|0\rangle+d|1\rangle)$. This setup works as follows: Suppose the input ququart state is $(a|H\rangle+b|V\rangle)\otimes(c|0\rangle+d|1\rangle)$.  After passing through QWP3 and HWP3, which are used to convert $a|H\rangle+b|V\rangle$ to $|H\rangle$, the ququart state becomes $|H\rangle\otimes(c|0\rangle+d|1\rangle)$. The subsequent two HWPs (one at $45^\circ$ and the other at $0^\circ$) and the BD are used to convert  $|H\rangle\otimes(c|0\rangle+d|1\rangle)$ to $c|H\rangle+d|V\rangle$, which is now a polarization qubit state. QWP4 and HWP4 are then used to convert  $c|H\rangle+d|V\rangle$ to $|H\rangle$, which can pass through the PBS and get detected by the SPD. As a result, for any input ququart state, only its $(a|H\rangle+b|V\rangle)\otimes(c|0\rangle+d|1\rangle)$ component can pass through the setup described above, which effectively realize the desired projective measurement. By changing the parameters a, b, c and d, this state analyzer can be used to perform a full state tomography on the input ququart state.

\begin{figure}
\includegraphics[width=0.5\textwidth]{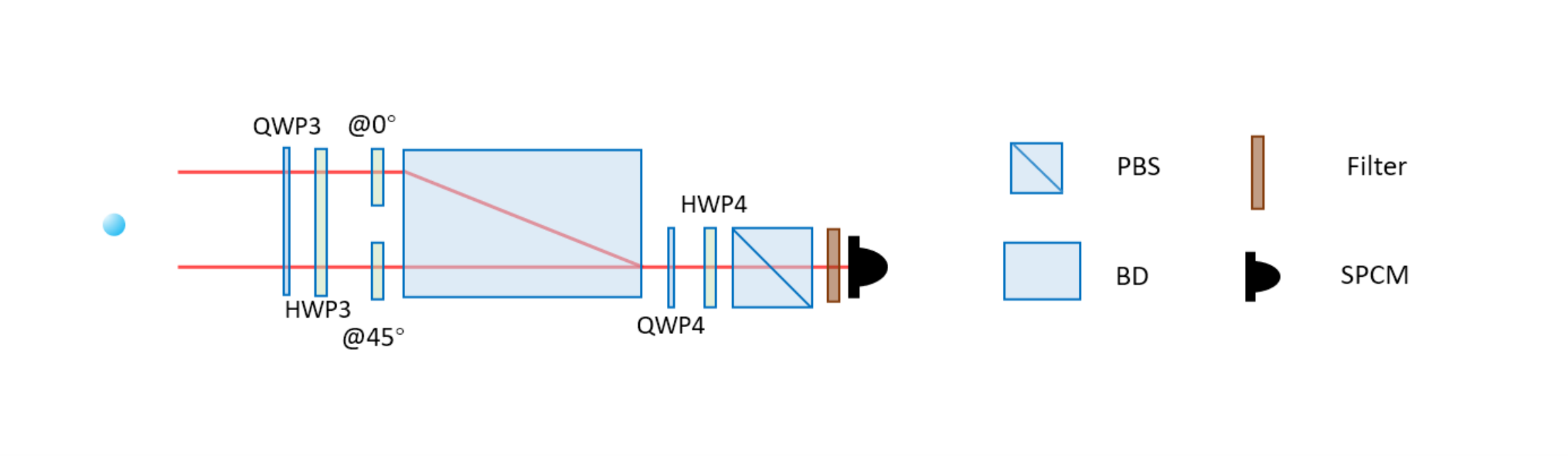}
\caption{State analyzer for a single-photon ququart state with both polarization and spatial degrees of freedom.}
\end{figure}

\textbf{E. Merge operation.}
The scheme of 2-to-4 QIT can be extended to the higher dimensional case, i.e., transferring one qubit of unknown quantum information from a qubit to a qudit. 
We call the operation of aggregating quantum information from two particles to one particle the Merge operation, and Merge(2, d$\rightarrow$2d) denotes the aggregation of quantum information of a qubit and a d-dimensional qudit to a 2d-dimensional qudit.
The quantum circuit to implement Merge(2, d$\rightarrow$2d) is shown in Fig. 10a, where the initial state of system A is $\alpha \vert  \bold{0}\rangle_A+\beta \vert  \bold{1}\rangle_A$ and the initial state of system B is $\sum_{\bold{i}=0}^{d-1}\gamma_i|\bold{i}\rangle$, which is a d-dimensional qudit. A controlled-$X_{2d}$~(C$X_{2d}$) gate is applied to A and B, where $X_{2d}$ is a 2d-dimensional unitary gate defined as $X_{2d}=\sum_{\bold{k}=0}^{d-1}(|\bold{k}\rangle\langle \bold{k+d}|+|\bold{k+d}\rangle\langle \bold{k}|)$, which swaps $|\bold{k}\rangle$ and $|\bold{k+d}\rangle$ for k in the range of 0 to d-1, expanding the state space of system B from d to 2d dimensions. The state of  A and B is thus converted from
\begin{equation}
\begin{split}
&(\alpha|\bold{0}\rangle+\beta|\bold{1}\rangle)\otimes\sum_{i=0}^{d-1}\gamma_i|\bold{i}\rangle\\
=&\sum_{i=0}^{d-1}(\alpha\gamma_i|\bold{0}\rangle_A|\bold{i}\rangle_B+\beta\gamma_i|\bold{1}\rangle_A|\bold{i}\rangle_B)
\end{split}
\end{equation}
to
\[
\begin{split}
&\sum_{i=0}^{d-1}(\alpha\gamma_i|\bold{0}\rangle_A|\bold{i}\rangle_B+\beta\gamma_i|\bold{1}\rangle_A|\bold{i+d}\rangle_B)\\
=&|+\rangle_A\otimes\sum_{i=0}^{d-1}(\alpha\gamma_i|\bold{i}\rangle_B+\beta\gamma_i|\bold{i+d}\rangle_B)\\
+&|-\rangle_B\otimes\sum_{i=0}^{d-1}(\alpha\gamma_i|\bold{i}\rangle_B-\beta\gamma_i|\bold{i+d}\rangle_B).
\end{split}
\]
After measuring A in the $\vert\pm\rangle$ basis and forwarding the outcome to B, a 2d-dimensional unitary operation ($I_{2d}$ or $Z_{2d}$) is applied on B conditioned on the outcome and thus converts its state to
\begin{equation}
\sum_{i=0}^{d-1}(\alpha\gamma_i|\bold{i}\rangle_B+\beta\gamma_i|\bold{i+d}\rangle_B),
\end{equation}
where 
\[
Z_{2d}=I_{2d}-2\sum_{k=0}^{d-1}|\bold{k+d}\rangle\langle \bold{k+d}|.
\]
By comparing Eq. (6) and Eq. (5), one can see that the two quantum states have exactly the same form except for the difference in the state basis, which means that the quantum information originally stored in A and B has been merged into B.

\begin{figure*}
\includegraphics[width=0.9\textwidth]{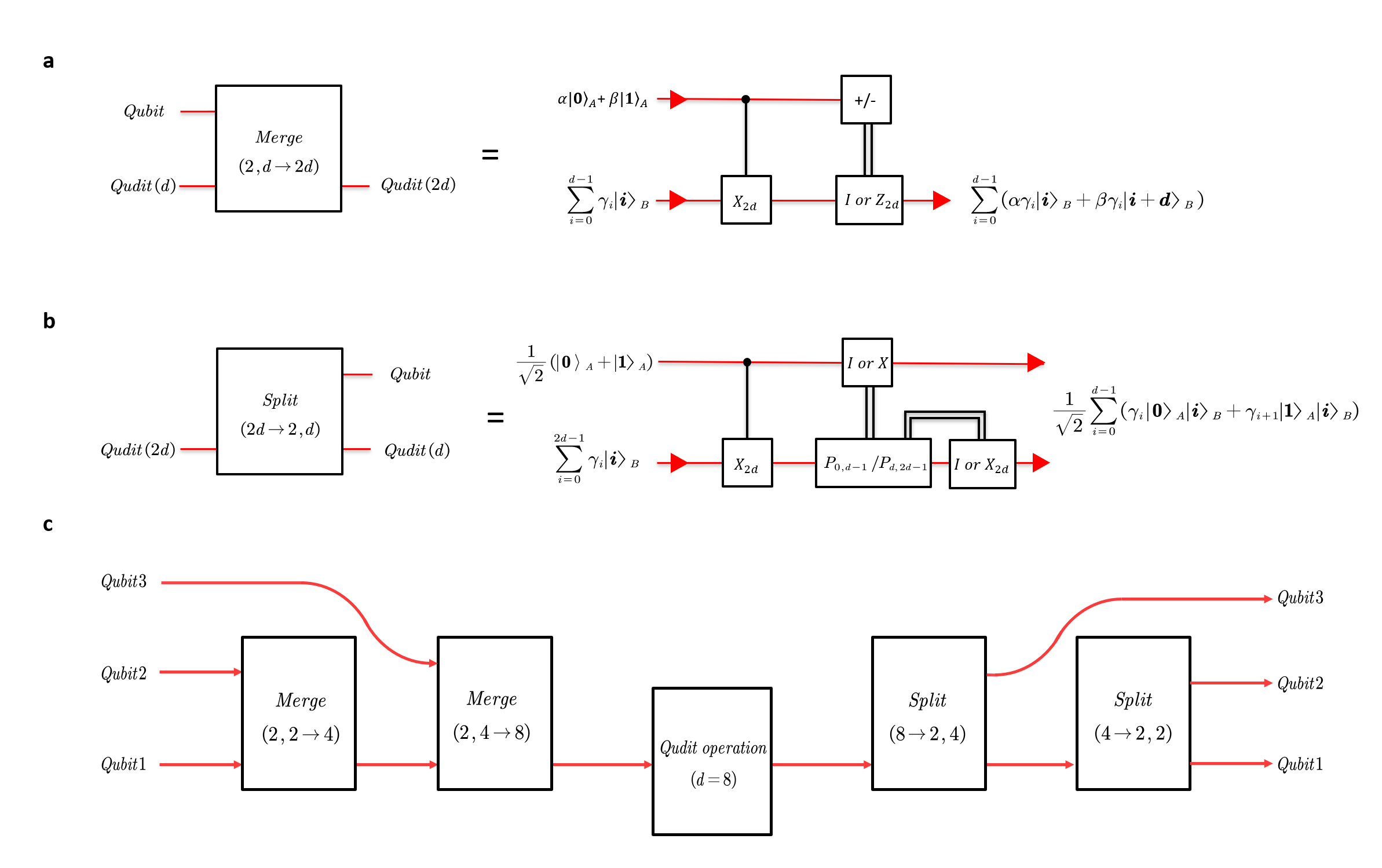}
\caption{$\bold{a,}$ The Merge operation. The quantum circuit for merging the quantum information of a qubit and a d-dimensional qudit into a 2d-dimensional qudit. $\bold{b,}$ The Split operation. The quantum circuit for splitting the quantum information of a 2d-dimensional qudit to a qubit and a d-dimensional qudit. $\bold{c,}$ Implementing a three-qubit quantum gate using Merge and Split operations.}\label{three-qubitgate}
\end{figure*}

\textbf{F. Split operation.}
The scheme of 4-to-2 QIT can be extended to the higher dimensional case, i.e., transferring one qubit of unknown quantum information from a qudit to a qubit. 
We call this operation of distributing quantum information from one particle to two particles the Split operation, and 
 Split(2d$\rightarrow$2, d) denotes the distribution of quantum information from a 2d-dimensional qudit to a qubit and a d-dimensional qudit. The quantum circuit to implement Split(2d$\rightarrow$2, d) is shown in Fig. 10b, where system A is initially in the state $\frac{1}{\sqrt{2}}(\vert  \bold{0}\rangle_A+\vert  \bold{1}\rangle_A)$ and the initial state of system B is in the 2d-dimensional qudit state
 \begin{equation}
 \sum_{\bold{i}=0}^{2d-1}\gamma_i|\bold{i}\rangle=\sum_{i=0}^{d-1}(\gamma_i|\bold{i}\rangle+\gamma_{i+d}|\bold{i+d}\rangle).
 \end{equation}
A C$X_{2d}$ gate is applied to A and B and their state becomes
\[
\begin{split}
&\frac{1}{\sqrt{2}}|\bold{0}\rangle\otimes \sum_{i=0}^{d-1}(\gamma_i|\bold{i}\rangle+\gamma_{i+d}|\bold{i+d}\rangle)\\
+&\frac{1}{\sqrt{2}}|\bold{1}\rangle \otimes \sum_{i=0}^{d-1}(\gamma_i|\bold{i+d}\rangle+\gamma_{i+d}|\bold{i}\rangle)\\
=&\frac{1}{\sqrt{2}}\sum_{i=0}^{d-1}(\gamma_i|\bold{0}\rangle |\bold{i}\rangle+\gamma_{i+d}|\bold{1}\rangle |\bold{i}\rangle) \\
+& \frac{1}{\sqrt{2}}\sum_{i=0}^{d-1}(\gamma_i|\bold{1}\rangle |\bold{i+d}\rangle+\gamma_{i+d}|\bold{0}\rangle |\bold{i+d}\rangle).
\end{split}
\]
A projective measurement is then applied on B to measure whether it is in the subspace spanned by $\vert  \bold{0}\rangle_B$, $\vert  \bold{1}\rangle_B$, ..., $\vert  \bold{d-2}\rangle_B$, and $\vert  \bold{d-1}\rangle_B$ or the subspace spanned by $\vert  \bold{d}\rangle_B$, $\vert  \bold{d+1}\rangle_B$, ..., $\vert  \bold{2d-2}\rangle_B$, and $\vert  \bold{2d-1}\rangle_B$. Based on the measurement outcome, unitary operations ($I\otimes I_{2d}$ or $X\otimes X_{2d}$) are applied on A and B and their state becomes
\begin{equation}
\frac{1}{\sqrt{2}}\sum_{i=0}^{d-1}(\gamma_i|\bold{0}\rangle |\bold{i}\rangle+\gamma_{i+d}|\bold{1}\rangle |\bold{i}\rangle).
\end{equation}
Comparing Eq. (8) and Eq. (7), it can be seen that the two quantum states have exactly the same form except for the difference in the state basis, which means that the quantum information originally stored in B are now split into A and B.

\textbf{G. Construction of multi-qubit gates.}
In various quantum information applications, including quantum computation and quantum simulation, multi-qubit quantum gates are widely used. 
Theoretically, multi-qubit quantum gates can be decomposed into two-qubit CNOT gates and single-qubit quantum gates for implementation, but in practice, such decomposition can be quite complex and consumes a lot of resources experimentally.
Here we propose a method to simplify the implementation of multi-qubit quantum gates by using QIT methods.
This approach essentially transforms an arbitrary n-qubit quantum gate operation into a unitary transform on a $2^n$-dimensional qudit, which is simpler to implement in certain circumstances.
For example, for a path-encoded $2^n$-dimensional photon, when n is not very large, an arbitrary $2^n\times2^n$ qudit unitary transform can be readily implemented using the Reck scheme \cite{Reck,Clements}.
We present below our method using a 3-qubit quantum gate as an example.
Specifically, the method can be divided into three steps. 
(i) The quantum information of the three input qubits is converged to a single particle, which can be achieved by two Merge operations.
As shown in Fig. 10c, a ququart is obtained by a Merge(2, 2$\rightarrow$4) operation acting on qubit 2 and qubit 1.
Then a Merge(2, 4$\rightarrow$8) operation acts on qubit 3 and this ququart to obtain a qudit (d=8), which contains the quantum information of the input three qubits.
(ii) The qudit is then subjected to a $8\times8$ unitary transformation, which has the same mathematical form as the matrix of the three-qubit quantum gate expanded in the computational basis.
(iii) The quantum information of this qudit is then distributed to three particles, which can be achieved by two Split operations.
The 8-dimensional qudit is split by a Split(8$\rightarrow2, 4$) operation to get a qubit and a ququart, and this ququart is then split by a Split(4$\rightarrow2, 2$) operation to finally get three qubits at the output, thus completing the three-qubit quantum gate.
For an arbitrary n-qubit quantum gate, it is often simpler and more resource-efficient to use our method than the traditional decomposition of n-qubit quantum gates into CNOT gates and single-qubit quantum gates, as long as n is not very large, i.e., the $2^n$-dimensional qudit can be easily unitary-transformed.\\


\begin{thebibliography}{99}
	\bibitem{wootters}
    Wootters, W. and Zurek, W. A single quantum cannot be cloned. Nature \text{299}, 802-803 (1982).
	\bibitem{Gisin}
	V. Scarani, S. Iblisdir, N. Gisin, and A. Acin,	Quantum cloning, Rev. Mod. Phys. \text{77}, 1225 (2005).
	
	\bibitem{Bennett}
	Bennett, C. H. \emph{et al.} Teleporting an unknown quantum state via dual classical and einstein-podolsky-rosen channels. Phys. Rev. Lett. \textbf{70}, 1895-1899 (1993).
	\bibitem{Pirandola}
	Pirandola, S., Eisert, J., Weedbrook, C. et al. Advances in quantum teleportation. Nature Photonics \textbf{9}, 641-652 (2015).
	
	\bibitem{Ren}
	Ren, J.-G. emph{et al.} Ground-to-satellite quantum teleportation. Nature 549, 70-73 (2017)
	\bibitem{Ekert}
	Ekert, A. K. Quantum cryptography based on bell's theorem. Phys. Rev. Lett. \textbf{67}, 661-663 (1991)
	\bibitem{Ma}
	Ma, X., Herbst, T., Scheidl, T. et al. Quantum teleportation over 143 kilometres using active feed-forward. Nature \textbf{489}, 269-273 (2012).
	\bibitem{Kimble}
	Kimble, H. J. The quantum internet. Nature \textbf{453}, 1023 (2008).
	\bibitem{Simon}
	Simon, C. Towards a global quantum network. Nature Photonics \textbf{11}, 678-680 (2017)
	\bibitem{Raussendorf}
	Raussendorf, R., \emph{et al.}  A one-way quantum computer. Phys. Rev. Lett. \textbf{86} (22), 5188-5191 (2001).
	\bibitem{Raussendorf2}
	Raussendorf, R., \emph{et al.} Measurement-based quantum computation on cluster states. Phys. Rev. A \textbf{68} (2), 022312 (2003).
	\bibitem{Walther}
	Walther, P. \emph{et al.} Experimental one-way quantum computing. Nature \textbf{434} (10), 169-176 (2005).
	
	\bibitem{d-oneway}
	Reimer C. \emph{et al.} High-dimensional one-way quantum processing implemented on d-level cluster states. Nature Physics \textbf{15}(2), 148-153 (2019).
	\bibitem{Gottesman}
	Gottesman, D and  Chuang, I. L . Demonstrating the viability of universal quantum computation using teleportation and single-qubit operations. Nature \textbf{402}(6070), 390-392 (1999).
		\bibitem{Kandel}
	Kandel, Y. P. \emph{et al.} Coherent spin-state transfer via Heisenberg exchange. Nature \textbf{573}(10), 553-557 (2019).
	\bibitem{He}
	He, Y \emph{et al.} Quantum State Transfer from a Single Photon to a Distant Quantum-Dot Electron Spin. Phys. Rev. Lett. \textbf{119}, 060501 (2017).
	
	
	\bibitem{Bouwmeester}
	Bouwmeester, Dik, \emph{et al.} Experimental quantum teleportation. Nature \textbf{390} (6660), 575-579  (1997).
	\bibitem{Furusawa}
	Furusawa, Akira, \emph{et al.} Unconditional quantum teleportation. science \textbf{282} (5389), 706-709 (1998).
	
	\bibitem{Bao}
	Bao, Xiao-Hui, \emph{et al.} Quantum teleportation between remote atomic-ensemble quantum memories. Proceedings of the National Academy of Sciences \textbf{109} (50), 20347-20351 (2012).
	\bibitem{Sherson}
	Sherson, J., Krauter, H., Olsson, R. et al. Quantum teleportation between light and matter. Nature \textbf{443}, 557-560 (2006).
	\bibitem{Riebe}
	Riebe, Mark, \emph{et al.} Deterministic quantum teleportation with atoms. Nature \textbf{429} (6993), 734-737 (2004).
	\bibitem{Barrett}
	Barrett, M. D., \emph{et al.} Deterministic quantum teleportation of atomic qubits. Nature \textbf{429} (6993),737-739 (2004).
	\bibitem{Qiao}
	Qiao, H., \emph{et al.} Conditional teleportation of quantum-dot spin states. Nature communications \textbf{11} (1), 1-9 (2020).
	\bibitem{Pfaff}
	Pfaff, W., \emph{et al.} Unconditional quantum teleportation between distant solid-state quantum bits. Science \textbf{345} (6196), 532-535 (2014).
\bibitem{Opt}
	Fiaschi, N., Hensen, B., Wallucks, A. et al. Optomechanical quantum teleportation. Nature Photonics \textbf{15}, 817-821 (2021).
	\bibitem{Steffen}
	Steffen, Lars, \emph{et al.} Deterministic quantum teleportation with feed-forward in a solid state system. Nature \textbf{500} (7462), 319-322 (2013).
	\bibitem{Yao}
	Blok, M. S., et al. Quantum information scrambling on a superconducting qutrit processor. Phy. Rev. X \textbf{11}(2), 021010 (2021).
	
	\bibitem{Zhao}
	Zhao, Z., \emph{et al.} Experimental demonstration of five-photon entanglement and open-destination teleportation. Nature \textbf{430} (6995), 54-58 (2004).
	\bibitem{Barasinski}
	Barasinski, A., \emph{et al.} Demonstration of controlled quantum teleportation for discrete variables on linear optical devices. Phys. Rev. Lett. \textbf{122} (17), 170501 (2019). 
	\bibitem{Zhang}
	Zhang, Q., \emph{et al.} Experimental quantum teleportation of a two-qubit composite system. Nature Physics \textbf{2}(10), 678-682 (2006).
	\bibitem{Luo}
	Luo, Y.-H., et al. Quantum teleportation of physical qubits into logical code spaces. Proceedings of the National Academy of Sciences \textbf{118}, 36 (2021).
	\bibitem{Hu}
	Hu, X., \emph{et al.} Experimental multi-level quantum teleportation. arXiv preprint arXiv:1904.12249 (2019).
	\bibitem{Luo}
	Luo, Y., \emph{et al.} Quantum Teleportation in High Dimensions. Phys. Rev. Lett. \textbf{123} (7), 070505 (2019). 
	\bibitem{Wang}
	Wang, X., \emph{et al.} Quantum teleportation of multiple degrees of freedom of a single photon. Nature \textbf{518} (7540), 516-519 (2015).
	
	\bibitem{Nielsen}
	Nielsen, M. A. and Chuang, I. L. Quantum Computation and Quantum Information (Cambridge Univ. Press, 2000)
	\bibitem{Barenco}
	Barenco, A., \emph{et al.} Elementary gates for quantum computation. Phys. Rev. A \textbf{52} (5), 3457 (1995)
	\bibitem{Vidal}
	Vidal, G., \emph{et al.} Universal quantum circuit for two-qubit transformations with three controlled-NOT gates. Phys. Rev. A  \textbf{69} (1), 010301 (2004).
	\bibitem{O'Brien}
	O'Brien, J. L., \emph{et al.} Demonstration of an all-optical quantum controlled-NOT gate. Nature \textbf{426} (6964), 264-267 (2003).
	\bibitem{Chou}
	Chou, K. S., \emph{et al.} Deterministic teleportation of a quantum gate between two logical qubits. Nature \textbf{561} (7723), 368-373 (2018).
	\bibitem{Fedorov}
	Fedorov, A., \emph{et al.} Implementation of a Toffoli gate with superconducting circuits. Nature \textbf{481} (7380), 170-172 (2012).
	\bibitem{Reed}
	Reed, M. D., \emph{et al.} Realization of three-qubit quantum error correction with superconducting circuits. Nature \textbf{482} (7385), 382-385 (2012).
	
	\bibitem{CNOT} Okamoto, R., \emph{et al.}  Demonstration of an optical quantum controlled-NOT gate without path interference. Phys. Rev. Lett.  \textbf{95} (21), 210506. (2005).
	
	
	\bibitem{Massar}
	Massar, S.,  \emph{et al.} Optimal Extraction ofInformation from Finite Quantum Ensembles, Phys. Rev. Lett. \textbf{74} (8), 1259-1263 (1995).
	\bibitem{Ralph}
	Ralph, T. C., \emph{et al.} A. Efficient Toffoli gates using qudits. Phys. Rev. A \textbf{75}, 022313 (2007)
	\bibitem{Lanyon}
	Lanyon, B. P., \emph{et al.} Simplifying quantum logic using higher-dimensional Hilbert spaces. Nat. Physics \textbf{5}(2),134-140 (2009).
	
	



\bibitem{ZhongHS} Zhong, H. S.  \emph{et al.} 12-photon entanglement and scalable scattershot boson sampling with optimal entangled-photon pairs from parametric down-conversion. Phys. Rev. Lett. \textbf{121} (25), 250505 (2018).

\bibitem{Langford}
Langford, N. K. \emph{et al.} Demonstration of a simple entangling optical gate and its use in Bell-state analysis. Phys. Rev. Lett. \textbf{95} (21), 210504 (2005).

\bibitem{Kiesel}
Kiesel, N. \emph{et al.} Linear optics controlled-phase gate made simple.  Phys. Rev. Lett.  \textbf{95} (21), 210505 (2005).

\bibitem{Reck}
Reck, M.  \emph{et al.}  Experimental realization of any discrete unitary operator. Phys. Rev. Lett. \textbf{73} (1) 58-61 (1994).

\bibitem{Clements} Clements, W. R.  \emph{et al.} Optimal design for universal multiport interferometers. Optica \textbf{3}~(12), 1460 (2016).

\end{thebibliography}
\end{document}